\documentclass[aps,twocolumn,floatfix,showpacs,amsmath,amsfonts]{revtex4}

\usepackage{graphicx}
\usepackage{bm}
\usepackage{units}
\usepackage{color}

\begin{document}

\title{Magnetoexcitons break antiunitary symmetries}

\author{Frank Schweiner}
\author{J\"org Main}
\author{G\"unter Wunner}
\affiliation{Institut f\"ur Theoretische Physik 1, Universit\"at Stuttgart,
  70550 Stuttgart, Germany}
\date{\today}

\begin{abstract}
We show analytically and numerically
that the application of an external magnetic 
field to highly excited Rydberg excitons
breaks all antiunitary symmetries in the system.
Only by considering the complete 
valence band structure of a direct band gap cubic semiconductor,
the Hamiltonian of excitons leads to 
the statistics of a 
Gaussian unitary ensemble (GUE) without the need for
interactions with other quasi-particles like phonons.
Hence, we give theoretical evidence
for a spatially homogeneous system breaking all antiunitary symmetries.

%

\end{abstract}


\maketitle

For more than 100 years
one distinguishes in classical mechanics 
between two fundamentally different
types of motion: regular and chaotic motion.
Their appearance strongly depends on the
presence of underlying symmetries, which
are connected with constants of motion and reduce the
degrees of freedom in a given system.
If symmetries are broken, the classical dynamics 
often becomes nonintegrable and chaotic. However, 
since the description of chaos by trajectories and
Lyapunov exponents is not possible in quantum mechanics,
it has been unknown
for a long time how classical chaos manifests itself in
quantum mechanical spectra~\cite{QSC,QCI}. 

The Bohigas-Giannoni-Schmit conjecture~\cite{QC_1} suggests
that quantum systems
with few degrees of freedom and with a chaotic classical
limit can be described by random matrix theory~\cite{QSC_29,QSC_30} and 
thus show typical level spacings.
At the transition to quantum chaos,
the level spacing statistics will
change from Poissonian statistics to the statistics of
a Gaussian orthogonal ensemble (GOE) or a 
Gaussian unitary ensemble (GUE) 
as symmetry reduction leads to a correlation of levels
and hence to a strong suppression of crossings~\cite{QSC}.

To which of the two universality classes
a given system belongs is determined
by remaining antiunitary symmetries in the system.
While GOE statistics can be observed in many different systems
like, e.g.,
in atomic~\cite{QSC_19,QC_2} and molecular spectra~\cite{QSC_18},
for nuclei in external magnetic fields~\cite{QC_5,QSC_11,QSC_12,QSC_13},
microwaves~\cite{QSC_15,QSC_16,QSC_17},
impurities~\cite{QC_3}, and quantum wells~\cite{QC_4},
GUE statistics appears only if \emph{all} antiunitary
symmetries are broken~\cite{QC_1,QC_15}.
Thus, GUE statistics is observable only in very exotic systems
like microwave cavities with ferrite strips~\cite{QSC_27}
or billards in microwave resonators~\cite{QSC_26}
and graphene quantum dots~\cite{QC_9}.

There is no example for a system showing GUE statistics in
atomic physics. 
This is especially true for one of the prime examples
when studying quantum chaos: the highly
excited hydrogen atom in strong external fields.
Even though the applied magnetic field breaks 
time-reversal invariance, 
at least one antiunitary symmetry, e.g.,
time reversal and a certain parity, remains
and GOE statistics is observed~\cite{QSC,QC_3_11,GUE1}.


Excitons are fundamental quasi-particles in semiconductors,
which consist of an electron in the conduction band
and a positively charged hole in the valence
band.
Recently, T.~Kazimierczuk~\emph{et~al}~\cite{GRE} 
have shown in a remarkable 
high-resolution absorption experiment
an almost perfect hydrogen-like absorption series
for the yellow exciton in cuprous oxide $\left(\mathrm{Cu_{2}O}\right)$ 
up to a principal quantum number 
of $n=25$. 
This experiment has drawn new interest to the field of excitons 
experimentally and theoretically~\cite{28,80,75,100,74,50,QC,76,77,150,125}.

Since excitons in semiconductors are often treated as the hydrogen
analog of the solid state but also show substantial
deviations from this behavior due to the surrounding solid,
the question about their level spacing statistics
in external fields arises.
First experimental investigations of the level spacing statistics
in an external magnetic field give indications on
a breaking of antiunitary symmetries, which is, however, 
attributed to the interaction of excitons and phonons~\cite{QC}.

Very recently, we have shown that it
is indispensable to account for the complete valence band structure
of $\mathrm{Cu_{2}O}$ in a  
quantitative theory of excitons~\cite{100}
to explain the striking experimental findings
of a fine structure splitting and the observability of
$F$ excitons~\cite{28}.
We have also proven that the effect
of the valence band structure on the exciton
spectra is even more prominent when treating
excitons in external
fields~\cite{125}.

In this Letter we will now 
show that the simultaneous
presence of a cubic band structure and
external fields will break all antiunitary
symmetries in the exciton system without the 
need of phonons. This effect
is present in all direct band gap semiconductors with a
cubic valence band structure and not restricted to $\mathrm{Cu_{2}O}$.
We prove not only analytically that the antiunitary
symmetry known from the hydrogen
atom in external fields is broken 
in the case of excitons, but also, by solving the
Schr\"odinger equation in a
complete basis, that the nearest-neighbor
spacing distribution of exciton states reveals GUE statistics.
Thus, we give the first theoretical evidence
for a spatially homogeneous system which breaks
all antiunitary symmetries and demonstrate a
fundamental difference between atoms in 
vacuum and excitons.

Without external fields the Hamiltonian of excitons in direct band gap 
semiconductors reads~\cite{100}
\begin{equation}
H=E_{\mathrm{g}}-e^{2}/4\pi\varepsilon_{0}\varepsilon\left|\boldsymbol{r}_{e}-\boldsymbol{r}_{h}\right|
+H_{\mathrm{e}}\left(\boldsymbol{p}_{\mathrm{e}}\right)+
H_{\mathrm{h}}\left(\boldsymbol{p}_{\mathrm{\mathrm{h}}}\right)\label{eq:Hpeph}
\end{equation}
with the band gap energy $E_{\mathrm{g}}$, the Coulomb interaction
which is screened by the dielectric constant $\varepsilon$, and the kinetic energies
of electron and hole.
While the conduction band is almost
parabolic in many semiconductors 
and thus the kinetic energy of the electron 
can be described by the simple expression 
$H_{\mathrm{e}}\left(\boldsymbol{p}_{\mathrm{e}}\right)=\boldsymbol{p}_{\mathrm{e}}^{2}/2m_{\mathrm{e}}$
with the effective mass $m_{\mathrm{e}}$, the 
kinetic energy
of the hole in the case of three 
coupled valence bands is given by the more complex
expression~\cite{25,80,100}
\begin{eqnarray}
H_{\mathrm{h}}\left(\boldsymbol{p}_{\mathrm{h}}\right) & = & H_{\mathrm{so}}+\left(\gamma_{1}+4\gamma_{2}\right)\boldsymbol{p}_{\mathrm{h}}^{2}/2m_{0}\nonumber \\
 & - & 3\gamma_{2}\left(p_{\mathrm{h}1}^{2}\boldsymbol{I}_{1}^{2}+\mathrm{c.p.}\right)/\hbar^{2}m_{0}\nonumber \\
 & - & 6\gamma_{3}\left(\left\{ p_{\mathrm{h}1},p_{\mathrm{h}2}\right\} \left\{ \boldsymbol{I}_{1},\boldsymbol{I}_{2}\right\} +\mathrm{c.p.}\right)/\hbar^{2}m_{0}.\label{eq:Hh}
\end{eqnarray}
with $\left\{a,b\right\}=(ab+ba)/2$.
Here $\gamma_i$ denote the three Luttinger parameters,
$m_0$ the free electron mass and c.p.~cyclic permutation.
The threefold degenerate valence band is accounted for by the
quasi-spin $I=1$, which is a convenient abstraction to denote the three 
orbital Bloch functions $xy$, $yz$, and $zx$~\cite{25}.
The components of its matrices $\boldsymbol{I}_{i}$ are given by $I_{i,\,jk}=-i\hbar\varepsilon_{ijk}$~\cite{25,100}
with the Levi-Civita symbol $\varepsilon_{ijk}$.
Note that the expression for $H_{\mathrm{h}}\left(\boldsymbol{p}_{\mathrm{h}}\right)$
can be separated in two parts having spherical and cubic symmetry, respectively~\cite{7_11}.
The coefficients $\mu'$ and $\delta'$
of these parts can be expressed in terms of the three Luttinger parameters:
$\mu'=\left(6\gamma_{3}+4\gamma_{2}\right)/5\gamma'_{1}$ and
$\delta'=\left(\gamma_{3}-\gamma_{2}\right)/\gamma'_{1}$
with $\gamma'_{1}=\gamma_{1}+m_{0}/m_{\mathrm{e}}$~\cite{7_11,7,100}.
Finally, the spin-orbit coupling
$H_{\mathrm{so}}=2\Delta/3\left(1+\boldsymbol{I}\cdot\boldsymbol{S}_{\mathrm{h}}/\hbar^{2}\right)$
between $I$ and the hole spin $S_{\mathrm{h}}$ describes a 
splitting of the valence bands at the center of the Brillouin zone~\cite{7}.

Let us now consider the case with external fields being present.
Then the corresponding Hamiltonian is obtained
via the minimal substitution. 
We further introduce relative and center of mass coordinates
and set the position and momentum of the 
center of mass to zero~\cite{90,91}.
Then the complete Hamiltonian
of the relative motion reads~\cite{34,33,39,TOE,44}
\begin{eqnarray}
H & = & E_{\mathrm{g}}-e^{2}/4\pi\varepsilon_{0}\varepsilon\left|\boldsymbol{r}\right|+H_{B}+e\Phi\left(\boldsymbol{r}\right)\nonumber \\
  & + & H_{\mathrm{e}}\left(\boldsymbol{p}+e\boldsymbol{A}\left(\boldsymbol{r}\right)\right)+H_{\mathrm{h}}\left(-\boldsymbol{p}+e\boldsymbol{A}\left(\boldsymbol{r}\right)\right).\label{eq:H}
\end{eqnarray}
with the relative coordinate $\boldsymbol{r}=\boldsymbol{r}_{\mathrm{e}}-\boldsymbol{r}_{\mathrm{h}}$
and the relative momentum $\boldsymbol{p}=\left(\boldsymbol{p}_{\mathrm{e}}-\boldsymbol{p}_{\mathrm{h}}\right)/2$
of electron and hole~\cite{90,91}.

Here we use the vector potential $\boldsymbol{A}=\left(\boldsymbol{B}\times\boldsymbol{r}\right)/2$
of a constant magnetic field $B$ and the electrostatic potential
$\Phi\left(\boldsymbol{r}\right)=-\boldsymbol{F}\cdot\boldsymbol{r}$ of a constant electric field $F$.
The term $H_{B}$ describes the energy
of the spins in the magnetic field~\cite{25,44,44_12,33}.
In this Letter we want to show that the Hamiltonian~(\ref{eq:H})
breaks all antiunitary symmetries. Since the term $H_B$ as well as
the spin orbit interaction are invariant under the 
symmetry operations considered below, we will neglect them in the following.

Before we investigate the symmetry of $H$, we have
to note that the matrices $\boldsymbol{I}_{i}$
are not the standard
spin matrices $\boldsymbol{S}_{i}$ 
of spin one~\cite{Messiah2}. However, since these matrices
obey the commutation rules~\cite{25}
\begin{equation}
\left[\boldsymbol{I}_{i},\,\boldsymbol{I}_{j}\right]=
i\hbar\sum_{k=1}^{3}\varepsilon_{ijk}\boldsymbol{I}_{k},
\end{equation}
there must be a unitary transformation 
$\boldsymbol{U}$ such that $\boldsymbol{U}^{\dagger}
\boldsymbol{I}_{i}\boldsymbol{U}=\boldsymbol{S}_{i}$
holds.
This transformation matrix reads
\begin{equation}
\boldsymbol{U}=\frac{1}{\sqrt{2}}\left(\begin{array}{ccc}
-1 & 0 & 1\\
-i & 0 & -i\\
0 & \sqrt{2} & 0
\end{array}\right)
\end{equation}
and we will now use the matrices $\boldsymbol{S}_{i}$
instead of $\boldsymbol{I}_{i}$ in the following.

In the special case with vanishing 
Luttinger parameters $\gamma_2=\gamma_3=0$, the exciton
Hamiltonian~(\ref{eq:H})
is of the same form as the Hamiltonian of a hydrogen atom
in external fields. 
It is well known that for this Hamiltonian
there is still one antiunitary symmetry left, i.e., that it
is invariant under the combined symmetry 
of time inversion $K$ followed by a reflection $S_{\hat{\boldsymbol{n}}}$ 
at the specific plane spanned by both fields~\cite{QSC}.
This plane is determined by the normal vector 
\begin{equation}
\hat{\boldsymbol{n}}=\left(\boldsymbol{B}\times\boldsymbol{F}\right)/\left|\boldsymbol{B}\times\boldsymbol{F}\right|\label{eq:nvec}
\end{equation}
or $\hat{\boldsymbol{n}}\perp\boldsymbol{B}/B$ if $F=0$ holds.
Therefore, the hydrogen-like system shows GOE statistics in the chaotic regime.

As the hydrogen atom is spherically symmetric in the field-free case,
it makes no difference whether the magnetic field is
oriented in $z$ direction or not. 
However, in a semiconductor with $\delta'\neq 0$ the Hamiltonian
has cubic symmetry and the orientation
of the external fields with respect to the crystal axis
of the lattice becomes important.
Any rotation of the coordinate system with the aim 
of making the $z$ axis coincide with the direction of the magnetic
field will also rotate the cubic crystal lattice.
Hence, we will show that the only remaining antiunitary symmetry
mentioned above is broken for the exciton Hamiltonian
if the plane spanned by both fields
is \emph{not} identical to one of the symmetry planes
of the cubic lattice.
Even without an external electric field the symmetry is broken
if the magnetic field is not oriented in 
one of these symmetry planes.
Only if the plane spanned by both fields
is identical to one of the symmetry planes of the cubic lattice,
the antiunitary symmetry $KS_{\hat{\boldsymbol{n}}}$
with $\hat{\boldsymbol{n}}$ given by Eq.~(\ref{eq:nvec})
is present.

At first, we will show this analytically.
Under time inversion $K$ and reflections $S_{\hat{\boldsymbol{n}}}$
at a plane perpendicular to a normal vector $\hat{\boldsymbol{n}}$
the vectors of position $\boldsymbol{r}$, momentum $\boldsymbol{p}$
and spin $\boldsymbol{S}$ transform according to~\cite{Messiah2}
\begin{equation}
K\boldsymbol{r}K^{\dagger}=\boldsymbol{r},\quad 
K\boldsymbol{p}K^{\dagger}=-\boldsymbol{p},\quad 
K\boldsymbol{S}K^{\dagger}=-\boldsymbol{S},
\end{equation}
and
\begin{subequations}
\begin{eqnarray}
S_{\hat{\boldsymbol{n}}}\boldsymbol{r}S_{\hat{\boldsymbol{n}}}^{\dagger} & = & \boldsymbol{r}-2\hat{\boldsymbol{n}}\left(\hat{\boldsymbol{n}}\cdot\boldsymbol{r}\right),\\
S_{\hat{\boldsymbol{n}}}\boldsymbol{p}S_{\hat{\boldsymbol{n}}}^{\dagger} & = & \boldsymbol{p}-2\hat{\boldsymbol{n}}\left(\hat{\boldsymbol{n}}\cdot\boldsymbol{p}\right),\\
S_{\hat{\boldsymbol{n}}}\boldsymbol{S}S_{\hat{\boldsymbol{n}}}^{\dagger} & = & -\boldsymbol{S}+2\hat{\boldsymbol{n}}\left(\hat{\boldsymbol{n}}\cdot\boldsymbol{S}\right).
\end{eqnarray}
\end{subequations}
Let us denote the orientation of $\boldsymbol{B}$ and $\boldsymbol{F}$
in spherical coordinates via
$\boldsymbol{B}\left(\varphi,\,\vartheta\right)=B\left(\cos\varphi\sin\vartheta,\,\sin\varphi\sin\vartheta,\,\cos\vartheta\right)^{\mathrm{T}}$.

Possible orientations of the fields breaking the antiunitary
symmetry are then, e.g., $\boldsymbol{B}\left(0,\,0\right)$ and $\boldsymbol{F}\left(\pi/6,\,\pi/2\right)$,
$\boldsymbol{B}\left(0,\,\pi/6\right)$ and $\boldsymbol{F}\left(\pi/2,\,\pi/2\right)$ or 
$\boldsymbol{B}\left(\pi/6,\,\pi/6\right)$ and $\boldsymbol{F}=\boldsymbol{0}$.
In all of these cases the hydrogen-like part of the Hamiltonian~(\ref{eq:H}) is
invariant under $KS_{\hat{\boldsymbol{n}}}$ with $\hat{\boldsymbol{n}}$ given by Eq.~(\ref{eq:nvec}). However, other parts
of the Hamiltonian like $H_c=\left(p_{1}^{2}\boldsymbol{S}_{1}^{2}+\mathrm{c.p.}\right)$ [see Eq.~(\ref{eq:Hh})]
are not invariant. For example, for the case with $\boldsymbol{B}\left(0,\,0\right)$ and $\boldsymbol{F}\left(\pi/6,\,\pi/2\right)$,
we obtain 
\begin{eqnarray}
& & S_{\hat{\boldsymbol{n}}}KH_{c}K^{\dagger}S_{\hat{\boldsymbol{n}}}^{\dagger}-H_{c}\nonumber\\
& = & 1/8\left[2\sqrt{3}\left(\boldsymbol{S}_{2}^{2}-\boldsymbol{S}_{1}^{2}\right)p_{1}p_{2}\right.\nonumber\\
& + & 3\left(\boldsymbol{S}_{1}^{2}p_{2}^{2}+\boldsymbol{S}_{2}^{2}p_{1}^{2}\right)-3\left(\boldsymbol{S}_{1}^{2}p_{1}^{2}+\boldsymbol{S}_{2}^{2}p_{2}^{2}\right)\nonumber\\
& + & \left.\left\{\boldsymbol{S}_{1},\boldsymbol{S}_{2}\right\}\left(2\sqrt{3}\left(p_{2}^{2}-p_{1}^{2}\right)+12p_{1}p_{2}\right)\right]\neq 0
\end{eqnarray}
with $\hat{\boldsymbol{n}}=\left(-1/2,\,\sqrt{3}/2,\,0\right)^{\mathrm{T}}$.
Thus, the generalized time-reversal symmetry
of the hydrogen atom is broken
for excitons due to the cubic symmetry of the semiconductor.

\begin{figure}
\begin{centering}
\includegraphics[width=1.0\columnwidth]{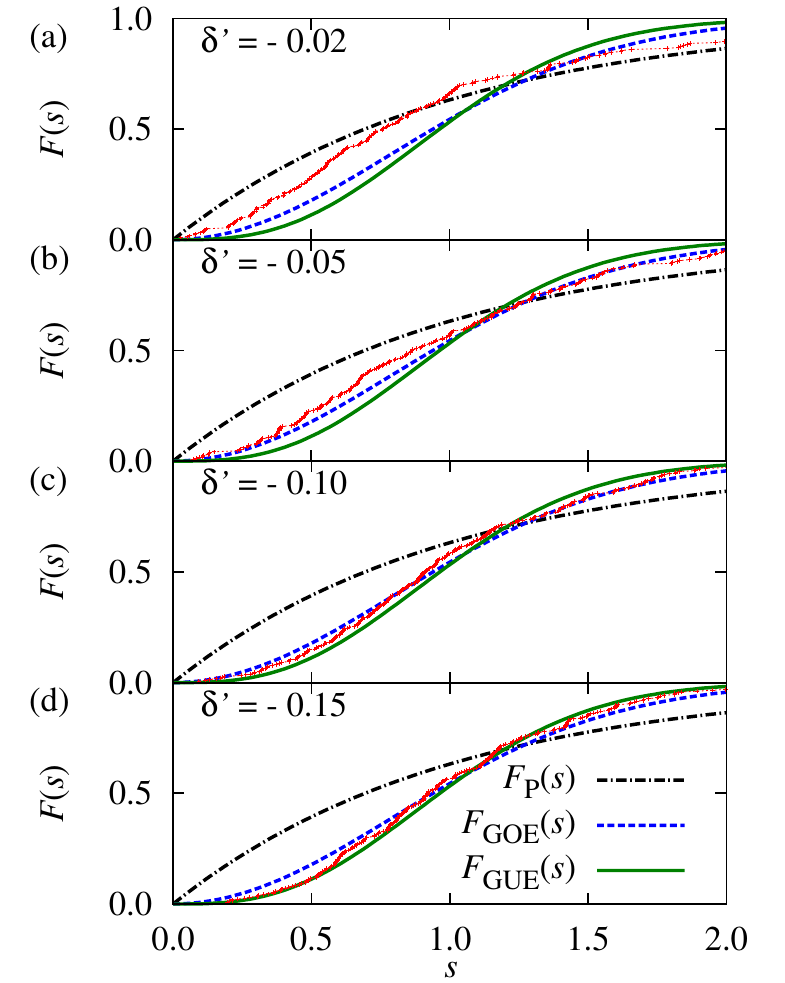}
\par\end{centering}

\protect\caption{Cumulative distribution function $F\left(s\right)$
for increasing values of $\delta'$ with $\boldsymbol{B}\left(\varphi,\,\vartheta\right)=\boldsymbol{B}\left(\pi/6,\,\pi/6\right)$ and $B=3\,\mathrm{T}$.
Besides the numerical data (red dots), we also show the corresponding functions of a 
Poissonian ensemble (black dash-dotted line), GOE (black dashed line),
and GUE (green solid line).
For increasing values of $\delta'$ the statistics rapidly changes to the
one of a Gaussian unitary ensemble (d).
Note that we do not show the hydrogen-like case
$\delta'=0$ since we simply obtain 
transitional form between Poissonian and GOE 
statistics and since this system is  
sufficiently well known from literature (see, e.g.,
Refs.~\cite{GUE1,QC_3_11,QSC} and further references therein).
~\label{fig:fig1}}
\end{figure}

Since a breaking of all antiunitary
symmetries is connected with the 
appearance of GUE statistics, we 
now solve the Schr\"odinger
equation corresponding to $H$ for the arbitrarily chosen
set of material parameters $E_{\mathrm{g}}=0$, $\varepsilon=7.5$, 
$m_{\mathrm{e}}=m_0$, $\gamma_1'=2$, and $\mu'=0$ using a complete basis. 
We can then analyze the nearest-neighbor spacings
of the energy levels~\cite{GUE1}.
To reduce the size of our basis and thus the numerical effort, 
we already assumed $\Delta=0$ so that we can disregard the
spins of electron and hole.

The cubic part of the Hamiltonian~(\ref{eq:H})
couples the angular momentum $L$ of the exciton and the
quasi spin $I$ to the total momentum $G=L+I$ with the
$z$ component $M_G$. 
For the radial part of the basis functions we use the Coulomb-Sturmian functions
of Refs.~\cite{100,S1} with the radial quantum number~$N$ to obtain a complete basis.
Hence the ansatz for the exciton wave function reads
\begin{eqnarray}
\left|\Psi\right\rangle  & = & \sum_{NLGM_G}c_{NLGM_G}\left|N,\,L,\,I,\,G,\,M_G\right\rangle,\label{eq:ansatz}
\end{eqnarray}
with complex coefficients $c$. 

Without an external electric field, 
parity is a good quantum number and
the operators in the Schr\"odinger equation
couple only basis states with even \emph{or} with odd
values of $L$. Hence, we consider the case with $\boldsymbol{B}\left(\pi/6,\,\pi/6\right)$ 
and $\boldsymbol{F}=\boldsymbol{0}$ and use only basis states with
odd values of $L$ as these exciton states can be observed in
direct band gap, parity forbidden semiconductors~\cite{100,28,74}.

After rotating the coordinate system by the Euler angles $\left(\alpha,\,\beta,\,\gamma\right)=\left(0,\,\vartheta,\,\varphi\right)$
to make the quantization axis coincide with the direction of the magnetic
field~\cite{44,ED}, we write the Hamiltonian in terms of irreducible tensors~\cite{ED,7_11}.
Inserting the ansatz~(\ref{eq:ansatz}) in the Schr\"odinger
equation $H\Psi=E\Psi$ and multiplying from the left with
the state $\left\langle N',\,L',\,I',\,G',\,M_G'\right|$,
we obtain a matrix representation 
of the Schr\"odinger equation of the form
$\boldsymbol{D}\boldsymbol{c}=E\boldsymbol{M}\boldsymbol{c}$.
The vector $\boldsymbol{c}$ contains the coefficients of the ansatz~(\ref{eq:ansatz})
and the matrix elements entering the matrices $\boldsymbol{D}$ and $\boldsymbol{M}$ can be
calculated using the relations given in Ref.~\cite{100}.
The generalized eigenvalue problem
is finally solved using an appropriate LAPACK routine~\cite{Lapack}.
In our numerical calculations, the maximum number of basis states used
is limited by the condition $N+L\leq 29$
due to the required computer memory.

\begin{figure}
\begin{centering}
\includegraphics[width=1.0\columnwidth]{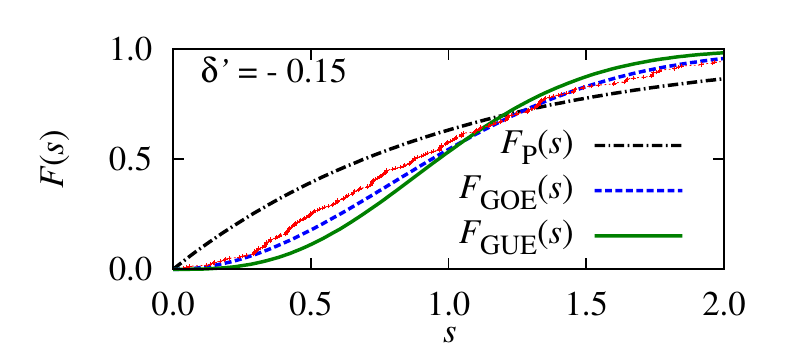}
\par\end{centering}

\protect\caption{Cumulative distribution function $F\left(s\right)$
for $\delta'=-0.15$ with $\boldsymbol{B}\left(\varphi,\,\vartheta\right)=\boldsymbol{B}\left(0,\,\pi/6\right)$ and $B=3\,\mathrm{T}$.
Since $\boldsymbol{B}$ is oriented in one of the symmetry planes of the lattice,
only GOE statistics can be observed when neglecting phonons.
~\label{fig:fig2}}
\end{figure}

Before analyzing the nearest-neighbor spacings,
we have to unfold the spectra to obtain 
a constant mean spacing~\cite{GUE1,QSC,QC_1,QC_16}.
The number of level spacings analyzed
is comparatively small since the magnetic field
breaks all symmetries in the system and impedes the
convergence of the solutions of the 
generalized eigenvalue problem with high energies~\cite{100}.
As in Ref.~\cite{GUE1}, we furthermore have to leave out
a certain number of low-lying sparse levels to remove individual but
nontypical fluctuations. Hence, we use
about $250$ exciton states for our analysis.
Owing to this number of states, we do not 
present histograms of the level
spacing probability distribution function $P(s)$
but calculate the cumulative distribution function~\cite{GUE2}
\begin{equation}
F(s)=\int_{0}^{s}P(x)\,\mathrm{d}x.
\end{equation}

The function $F(s)$ is shown in Fig.~\ref{fig:fig1} 
for increasing values of the parameter $\delta'$ at $B=3\,\mathrm{T}$.
In this figure we also show the
cumulative distribution function corresponding to the
spacing distributions known from random
matrix theory~\cite{QC_1,QC}:
the Poissonian distribution 
\begin{equation}
P_{\mathrm{P}}(s)=e^{-s}
\end{equation}
for non-interacting energy levels, the Wigner distribution 
\begin{equation}
P_{\mathrm{GOE}}(s)=\frac{\pi}{2}\,se^{-\pi s^2/4},
\end{equation} 
and the distribution 
\begin{equation}
P_{\mathrm{GUE}}(s)=\frac{32}{\pi^2}\,s^2e^{-4 s^2/\pi}
\end{equation}
for systems without any antiunitary symmetry. 
Note that the most
characteristic feature of GUE statistics is
the quadratic level repulsion for small
$s$ and that the clearest distinction
between GOE and GUE statistics can
be taken for $0\leq s\lesssim 0.5$.
Hence, we see from  that there is clear evidence for
GUE statistics.
Note that for all results presented in 
Fig.~\ref{fig:fig1} we used the constant value of $B=3\,\mathrm{T}$
and exciton states within a certain energy range. It is well known from
atomic physics that chaotic effects become more apparent
in higher magnetic fields or by using states of higher energies for the
analysis. Hence, by increasing $B$ or investigating the statistics
of exciton states with higher energies, GUE statistics could
probably be observed for smaller values of $|\delta'|$.
At this point we have to note that 
an evaluation of numerical spectra for $\delta'>0$ shows the same appearance of GUE statistics.
This is expected since the analytically shown breaking of all antiunitary symmetries
is independent of the sign of the material parameters.

If the magnetic field is oriented in one of the symmetry planes of the
cubic lattice, only GOE statistics is observable. Indeed, when 
investigating the exciton spectrum for, e.g., $\boldsymbol{B}\left(0,\,\pi/6\right)$,
the level spacing statistics is best described by GOE statistics, 
especially for small values of $s$, as can be seen from Fig.~\ref{fig:fig2}.
Very recently, M.~A{\ss}mann~\emph{et~al}~\cite{QC}
have shown experimentally that 
excitons in $\mathrm{Cu_{2}O}$ 
show GUE statistics in an external magnetic field.
However, since their experimental spectra
were analyzed exactly for $\boldsymbol{B}\left(0,\,\pi/6\right)$,
there must be another explanation for this
observation than the cubic band structure. M.~A{\ss}mann~\emph{et~al}~\cite{QC}
have assigned the observation of GUE statistics
to the interaction of
excitons and phonons.

The main advantage of theory over the experiments is the fact that the
exciton-phonon interaction can be left out.
Hence, one can treat the effects of the band structure and of the
exciton-phonon interaction separately. 
We performed model calculations to demonstrate that, in general,
GUE statistics appears for a much simpler system, i.e.,
only the presence of the cubic lattice
and the external fields already breaks 
all antiunitary symmetries
without the need for interactions with other
quasi-particles like phonons.
We did not intend a line-by-line comparison with experimental results.
Due to the high dimension of the problem as a result of the presence of the complex band structure, the spin orbit interaction,
and phonons, this is not possible at the moment.
However, we do not expect that the effects of the band structure and the phonons
on the level spacing statistics will cancel each other out. Indeed,
based on the analytic part of our analysis and the fact that the operator 
describing the interaction between excitons and phonons looks quite different
from the operators in our Hamiltonian~\cite{TOE},
the phonons certainly do not restore antiunitary symmetries.
Instead, the results of Ref.~\cite{QC} suggest that phonons will further increase the chaos.

We think that $\mathrm{Cu_{2}O}$ is the most promising candidate to investigate the
effect of the band structure. As the experiments in Ref.~\cite{QC} were performed with the magnetic
field being oriented in a direction of high symmetry, it would now be highly desirable
to investigate exciton absorption spectra in $\mathrm{Cu_{2}O}$ for other orientations of
the magnetic field to observe the effect of the band structure on the line statistics.

In conclusion, we have shown analytically and
numerically that the
cubic symmetry of the lattice and the band structure
leads to a breaking of all 
antiunitary symmetries in the system of magnetoexcitons. This effect 
demonstrates a fundamental difference between atoms in 
vacuum and excitons and is not limited to certain
values of the material parameters, for which reason it appears in
all direct band gap semiconductors with a cubic valence band structure. 
Furthermore, a closer investigation of excitons in external fields
can lead to a better understanding of the connection
between quantum and classical chaos.

We thank D. Fr\"ohlich for helpful discussions.



\begin{thebibliography}{52}
\expandafter\ifx\csname natexlab\endcsname\relax\def\natexlab#1{#1}\fi
\expandafter\ifx\csname bibnamefont\endcsname\relax
  \def\bibnamefont#1{#1}\fi
\expandafter\ifx\csname bibfnamefont\endcsname\relax
  \def\bibfnamefont#1{#1}\fi
\expandafter\ifx\csname citenamefont\endcsname\relax
  \def\citenamefont#1{#1}\fi
\expandafter\ifx\csname url\endcsname\relax
  \def\url#1{\texttt{#1}}\fi
\expandafter\ifx\csname urlprefix\endcsname\relax\def\urlprefix{URL }\fi
\providecommand{\bibinfo}[2]{#2}
\providecommand{\eprint}[2][]{\url{#2}}

\bibitem[{\citenamefont{Haake}(2010)}]{QSC}
\bibinfo{author}{\bibfnamefont{F.}~\bibnamefont{Haake}},
  \emph{\bibinfo{title}{Quantum Signatures of Chaos}}, Springer Series in
  Synergetics (\bibinfo{publisher}{Springer}, \bibinfo{address}{Heidelberg},
  \bibinfo{year}{2010}), \bibinfo{edition}{3rd} ed.

\bibitem[{\citenamefont{St\"{o}ckmann}(1999)}]{QCI}
\bibinfo{author}{\bibfnamefont{H.-J.} \bibnamefont{St\"{o}ckmann}},
  \emph{\bibinfo{title}{Quantum Chaos: An Introduction}}
  (\bibinfo{publisher}{Cambridge University Press},
  \bibinfo{address}{Cambridge}, \bibinfo{year}{1999}).

\bibitem[{\citenamefont{Bohigas et~al.}(1984)\citenamefont{Bohigas, Giannoni,
  and Schmit}}]{QC_1}
\bibinfo{author}{\bibfnamefont{O.}~\bibnamefont{Bohigas}},
  \bibinfo{author}{\bibfnamefont{M.~J.} \bibnamefont{Giannoni}},
  \bibnamefont{and} \bibinfo{author}{\bibfnamefont{C.}~\bibnamefont{Schmit}},
  \bibinfo{journal}{Phys. Rev. Lett.} \textbf{\bibinfo{volume}{52}},
  \bibinfo{pages}{1} (\bibinfo{year}{1984}).

\bibitem[{\citenamefont{Mehta}(2004)}]{QSC_29}
\bibinfo{author}{\bibfnamefont{M.~L.} \bibnamefont{Mehta}},
  \emph{\bibinfo{title}{Random Matrices}} (\bibinfo{publisher}{Elsevier},
  \bibinfo{address}{Amsterdam}, \bibinfo{year}{2004}), \bibinfo{edition}{3rd}
  ed.

\bibitem[{\citenamefont{Porter}(1965)}]{QSC_30}
\bibinfo{editor}{\bibfnamefont{C.~E.} \bibnamefont{Porter}}, ed.,
  \emph{\bibinfo{title}{Statistical Theory of Spectra}}
  (\bibinfo{publisher}{Academic Press}, \bibinfo{address}{New York},
  \bibinfo{year}{1965}).

\bibitem[{\citenamefont{Held et~al.}(1998)\citenamefont{Held, Schlichter,
  Raithel, and Walther}}]{QSC_19}
\bibinfo{author}{\bibfnamefont{H.}~\bibnamefont{Held}},
  \bibinfo{author}{\bibfnamefont{J.}~\bibnamefont{Schlichter}},
  \bibinfo{author}{\bibfnamefont{G.}~\bibnamefont{Raithel}}, \bibnamefont{and}
  \bibinfo{author}{\bibfnamefont{H.}~\bibnamefont{Walther}},
  \bibinfo{journal}{Europhys. Lett.} \textbf{\bibinfo{volume}{43}},
  \bibinfo{pages}{392} (\bibinfo{year}{1998}).

\bibitem[{\citenamefont{Frisch et~al.}(2014)\citenamefont{Frisch, Mark, Aikawa,
  Ferlaino, Bohn, Makrides, Petrov, and Kotochigova}}]{QC_2}
\bibinfo{author}{\bibfnamefont{A.}~\bibnamefont{Frisch}},
  \bibinfo{author}{\bibfnamefont{M.}~\bibnamefont{Mark}},
  \bibinfo{author}{\bibfnamefont{K.}~\bibnamefont{Aikawa}},
  \bibinfo{author}{\bibfnamefont{F.}~\bibnamefont{Ferlaino}},
  \bibinfo{author}{\bibfnamefont{J.~L.} \bibnamefont{Bohn}},
  \bibinfo{author}{\bibfnamefont{C.}~\bibnamefont{Makrides}},
  \bibinfo{author}{\bibfnamefont{A.}~\bibnamefont{Petrov}}, \bibnamefont{and}
  \bibinfo{author}{\bibfnamefont{S.}~\bibnamefont{Kotochigova}},
  \bibinfo{journal}{Nature} \textbf{\bibinfo{volume}{507}},
  \bibinfo{pages}{475} (\bibinfo{year}{2014}).

\bibitem[{\citenamefont{Zimmermann et~al.}(1988)\citenamefont{Zimmermann,
  K\"{o}ppel, Cederbaum, Persch, and Demtr\"{o}der}}]{QSC_18}
\bibinfo{author}{\bibfnamefont{T.}~\bibnamefont{Zimmermann}},
  \bibinfo{author}{\bibfnamefont{H.}~\bibnamefont{K\"{o}ppel}},
  \bibinfo{author}{\bibfnamefont{L.~S.} \bibnamefont{Cederbaum}},
  \bibinfo{author}{\bibfnamefont{G.}~\bibnamefont{Persch}}, \bibnamefont{and}
  \bibinfo{author}{\bibfnamefont{W.}~\bibnamefont{Demtr\"{o}der}},
  \bibinfo{journal}{Phys. Rev. Lett.} \textbf{\bibinfo{volume}{61}},
  \bibinfo{pages}{3} (\bibinfo{year}{1988}).

\bibitem[{\citenamefont{Mitchell et~al.}(2010)\citenamefont{Mitchell, Richter,
  and Weidenm\"{u}ller}}]{QC_5}
\bibinfo{author}{\bibfnamefont{G.~E.} \bibnamefont{Mitchell}},
  \bibinfo{author}{\bibfnamefont{A.}~\bibnamefont{Richter}}, \bibnamefont{and}
  \bibinfo{author}{\bibfnamefont{H.~A.} \bibnamefont{Weidenm\"{u}ller}},
  \bibinfo{journal}{Rev. Mod. Phys.} \textbf{\bibinfo{volume}{82}},
  \bibinfo{pages}{2845} (\bibinfo{year}{2010}).

\bibitem[{\citenamefont{Brody et~al.}(1981{\natexlab{a}})\citenamefont{Brody,
  Flores, French, Mello, Pandey, and Wong}}]{QSC_11}
\bibinfo{author}{\bibfnamefont{T.~A.} \bibnamefont{Brody}},
  \bibinfo{author}{\bibfnamefont{J.}~\bibnamefont{Flores}},
  \bibinfo{author}{\bibfnamefont{J.~B.} \bibnamefont{French}},
  \bibinfo{author}{\bibfnamefont{P.~A.} \bibnamefont{Mello}},
  \bibinfo{author}{\bibfnamefont{A.}~\bibnamefont{Pandey}}, \bibnamefont{and}
  \bibinfo{author}{\bibfnamefont{S.~S.~M.} \bibnamefont{Wong}},
  \bibinfo{journal}{Rev. Mod. Phys.} \textbf{\bibinfo{volume}{53}},
  \bibinfo{pages}{385} (\bibinfo{year}{1981}{\natexlab{a}}).

\bibitem[{\citenamefont{Rosenzweig and Porter}(1960)}]{QSC_12}
\bibinfo{author}{\bibfnamefont{N.}~\bibnamefont{Rosenzweig}} \bibnamefont{and}
  \bibinfo{author}{\bibfnamefont{C.~E.} \bibnamefont{Porter}},
  \bibinfo{journal}{Phys. Rev.} \textbf{\bibinfo{volume}{120}},
  \bibinfo{pages}{1698} (\bibinfo{year}{1960}).

\bibitem[{\citenamefont{Camarda and Georgopulos}(1983)}]{QSC_13}
\bibinfo{author}{\bibfnamefont{H.~S.} \bibnamefont{Camarda}} \bibnamefont{and}
  \bibinfo{author}{\bibfnamefont{P.~D.} \bibnamefont{Georgopulos}},
  \bibinfo{journal}{Phys. Rev. Lett.} \textbf{\bibinfo{volume}{50}},
  \bibinfo{pages}{492} (\bibinfo{year}{1983}).

\bibitem[{\citenamefont{St\"{o}ckmann and Stein}(1990)}]{QSC_15}
\bibinfo{author}{\bibfnamefont{H.-J.} \bibnamefont{St\"{o}ckmann}}
  \bibnamefont{and} \bibinfo{author}{\bibfnamefont{J.}~\bibnamefont{Stein}},
  \bibinfo{journal}{Phys. Rev. Lett.} \textbf{\bibinfo{volume}{64}},
  \bibinfo{pages}{2215} (\bibinfo{year}{1990}).

\bibitem[{\citenamefont{Alt et~al.}(1995)\citenamefont{Alt, Gr\"{a}f, Harney,
  Hofferbert, Lengeler, Richter, Schardt, and Weidenm\"{u}ller}}]{QSC_16}
\bibinfo{author}{\bibfnamefont{H.}~\bibnamefont{Alt}},
  \bibinfo{author}{\bibfnamefont{H.-D.} \bibnamefont{Gr\"{a}f}},
  \bibinfo{author}{\bibfnamefont{H.~L.} \bibnamefont{Harney}},
  \bibinfo{author}{\bibfnamefont{R.}~\bibnamefont{Hofferbert}},
  \bibinfo{author}{\bibfnamefont{H.}~\bibnamefont{Lengeler}},
  \bibinfo{author}{\bibfnamefont{A.}~\bibnamefont{Richter}},
  \bibinfo{author}{\bibfnamefont{P.}~\bibnamefont{Schardt}}, \bibnamefont{and}
  \bibinfo{author}{\bibfnamefont{H.~A.} \bibnamefont{Weidenm\"{u}ller}},
  \bibinfo{journal}{Phys. Rev. Lett.} \textbf{\bibinfo{volume}{74}},
  \bibinfo{pages}{62} (\bibinfo{year}{1995}).

\bibitem[{\citenamefont{Alt et~al.}(1996)\citenamefont{Alt, Gr\"{a}f,
  Hofferbert, Rangacharyulu, Rehfeld, Richter, Schardt, and Wirzba}}]{QSC_17}
\bibinfo{author}{\bibfnamefont{H.}~\bibnamefont{Alt}},
  \bibinfo{author}{\bibfnamefont{H.-D.} \bibnamefont{Gr\"{a}f}},
  \bibinfo{author}{\bibfnamefont{R.}~\bibnamefont{Hofferbert}},
  \bibinfo{author}{\bibfnamefont{C.}~\bibnamefont{Rangacharyulu}},
  \bibinfo{author}{\bibfnamefont{H.}~\bibnamefont{Rehfeld}},
  \bibinfo{author}{\bibfnamefont{A.}~\bibnamefont{Richter}},
  \bibinfo{author}{\bibfnamefont{P.}~\bibnamefont{Schardt}}, \bibnamefont{and}
  \bibinfo{author}{\bibfnamefont{A.}~\bibnamefont{Wirzba}},
  \bibinfo{journal}{Phys. Rev. E} \textbf{\bibinfo{volume}{54}},
  \bibinfo{pages}{2303} (\bibinfo{year}{1996}).

\bibitem[{\citenamefont{Zhou et~al.}(2010)\citenamefont{Zhou, Chen, Zhang, Yu,
  Lu, and Shen}}]{QC_3}
\bibinfo{author}{\bibfnamefont{W.}~\bibnamefont{Zhou}},
  \bibinfo{author}{\bibfnamefont{Z.}~\bibnamefont{Chen}},
  \bibinfo{author}{\bibfnamefont{B.}~\bibnamefont{Zhang}},
  \bibinfo{author}{\bibfnamefont{C.~H.} \bibnamefont{Yu}},
  \bibinfo{author}{\bibfnamefont{W.}~\bibnamefont{Lu}}, \bibnamefont{and}
  \bibinfo{author}{\bibfnamefont{S.~C.} \bibnamefont{Shen}},
  \bibinfo{journal}{Phys. Rev. Lett.} \textbf{\bibinfo{volume}{105}},
  \bibinfo{pages}{024101} (\bibinfo{year}{2010}).

\bibitem[{\citenamefont{Vina et~al.}(1998)\citenamefont{Vina, Potemski, and
  Wang}}]{QC_4}
\bibinfo{author}{\bibfnamefont{L.}~\bibnamefont{Vina}},
  \bibinfo{author}{\bibfnamefont{M.}~\bibnamefont{Potemski}}, \bibnamefont{and}
  \bibinfo{author}{\bibfnamefont{W.}~\bibnamefont{Wang}},
  \bibinfo{journal}{Phys.-Usp.} \textbf{\bibinfo{volume}{41}},
  \bibinfo{pages}{153} (\bibinfo{year}{1998}).

\bibitem[{\citenamefont{Seligman and Verbaarschot}(1985)}]{QC_15}
\bibinfo{author}{\bibfnamefont{T.}~\bibnamefont{Seligman}} \bibnamefont{and}
  \bibinfo{author}{\bibfnamefont{J.}~\bibnamefont{Verbaarschot}},
  \bibinfo{journal}{Phys. Lett. A} \textbf{\bibinfo{volume}{108}},
  \bibinfo{pages}{183} (\bibinfo{year}{1985}).

\bibitem[{\citenamefont{So et~al.}(1995)\citenamefont{So, Anlage, Ott, and
  Oerter}}]{QSC_27}
\bibinfo{author}{\bibfnamefont{P.}~\bibnamefont{So}},
  \bibinfo{author}{\bibfnamefont{S.~M.} \bibnamefont{Anlage}},
  \bibinfo{author}{\bibfnamefont{E.}~\bibnamefont{Ott}}, \bibnamefont{and}
  \bibinfo{author}{\bibfnamefont{R.~N.} \bibnamefont{Oerter}},
  \bibinfo{journal}{Phys. Rev. Lett.} \textbf{\bibinfo{volume}{74}},
  \bibinfo{pages}{2662} (\bibinfo{year}{1995}).

\bibitem[{\citenamefont{Stoffregen et~al.}(1995)\citenamefont{Stoffregen,
  Stein, St\"{o}ckmann, Ku\'{s}, and Haake}}]{QSC_26}
\bibinfo{author}{\bibfnamefont{U.}~\bibnamefont{Stoffregen}},
  \bibinfo{author}{\bibfnamefont{J.}~\bibnamefont{Stein}},
  \bibinfo{author}{\bibfnamefont{H.-J.} \bibnamefont{St\"{o}ckmann}},
  \bibinfo{author}{\bibfnamefont{M.}~\bibnamefont{Ku\'{s}}}, \bibnamefont{and}
  \bibinfo{author}{\bibfnamefont{F.}~\bibnamefont{Haake}},
  \bibinfo{journal}{Phys. Rev. Lett.} \textbf{\bibinfo{volume}{74}},
  \bibinfo{pages}{2666} (\bibinfo{year}{1995}).

\bibitem[{\citenamefont{Ponomarenko et~al.}(2008)\citenamefont{Ponomarenko,
  Schedin, Katsnelson, Yang, Hill, Novoselov, and Geim}}]{QC_9}
\bibinfo{author}{\bibfnamefont{L.~A.} \bibnamefont{Ponomarenko}},
  \bibinfo{author}{\bibfnamefont{F.}~\bibnamefont{Schedin}},
  \bibinfo{author}{\bibfnamefont{M.~I.} \bibnamefont{Katsnelson}},
  \bibinfo{author}{\bibfnamefont{R.}~\bibnamefont{Yang}},
  \bibinfo{author}{\bibfnamefont{E.~W.} \bibnamefont{Hill}},
  \bibinfo{author}{\bibfnamefont{K.~S.} \bibnamefont{Novoselov}},
  \bibnamefont{and} \bibinfo{author}{\bibfnamefont{A.~K.} \bibnamefont{Geim}},
  \bibinfo{journal}{Science} \textbf{\bibinfo{volume}{320}},
  \bibinfo{pages}{356} (\bibinfo{year}{2008}).

\bibitem[{\citenamefont{Friedrich and Wintgen}(1989)}]{QC_3_11}
\bibinfo{author}{\bibfnamefont{H.}~\bibnamefont{Friedrich}} \bibnamefont{and}
  \bibinfo{author}{\bibfnamefont{D.}~\bibnamefont{Wintgen}},
  \bibinfo{journal}{Phys. Rep.} \textbf{\bibinfo{volume}{183}},
  \bibinfo{pages}{37} (\bibinfo{year}{1989}).

\bibitem[{\citenamefont{Wintgen and Friedrich}(1987)}]{GUE1}
\bibinfo{author}{\bibfnamefont{D.}~\bibnamefont{Wintgen}} \bibnamefont{and}
  \bibinfo{author}{\bibfnamefont{H.}~\bibnamefont{Friedrich}},
  \bibinfo{journal}{Phys. Rev. A} \textbf{\bibinfo{volume}{35}},
  \bibinfo{pages}{1464(R)} (\bibinfo{year}{1987}).

\bibitem[{\citenamefont{Kazimierczuk et~al.}(2014)\citenamefont{Kazimierczuk,
  Fr\"{o}hlich, Scheel, Stolz, and Bayer}}]{GRE}
\bibinfo{author}{\bibfnamefont{T.}~\bibnamefont{Kazimierczuk}},
  \bibinfo{author}{\bibfnamefont{D.}~\bibnamefont{Fr\"{o}hlich}},
  \bibinfo{author}{\bibfnamefont{S.}~\bibnamefont{Scheel}},
  \bibinfo{author}{\bibfnamefont{H.}~\bibnamefont{Stolz}}, \bibnamefont{and}
  \bibinfo{author}{\bibfnamefont{M.}~\bibnamefont{Bayer}},
  \bibinfo{journal}{Nature} \textbf{\bibinfo{volume}{514}},
  \bibinfo{pages}{343} (\bibinfo{year}{2014}).

\bibitem[{\citenamefont{Thewes et~al.}(2015)\citenamefont{Thewes,
  Heck\"{o}tter, Kazimierczuk, A{\ss}mann, Fr\"{o}hlich, Bayer, Semina, and
  Glazov}}]{28}
\bibinfo{author}{\bibfnamefont{J.}~\bibnamefont{Thewes}},
  \bibinfo{author}{\bibfnamefont{J.}~\bibnamefont{Heck\"{o}tter}},
  \bibinfo{author}{\bibfnamefont{T.}~\bibnamefont{Kazimierczuk}},
  \bibinfo{author}{\bibfnamefont{M.}~\bibnamefont{A{\ss}mann}},
  \bibinfo{author}{\bibfnamefont{D.}~\bibnamefont{Fr\"{o}hlich}},
  \bibinfo{author}{\bibfnamefont{M.}~\bibnamefont{Bayer}},
  \bibinfo{author}{\bibfnamefont{M.~A.} \bibnamefont{Semina}},
  \bibnamefont{and} \bibinfo{author}{\bibfnamefont{M.~M.}
  \bibnamefont{Glazov}}, \bibinfo{journal}{Phys. Rev. Lett.}
  \textbf{\bibinfo{volume}{115}}, \bibinfo{pages}{027402}
  (\bibinfo{year}{2015}), \bibinfo{note}{and Supplementary Material}.

\bibitem[{\citenamefont{Sch\"{o}ne et~al.}(2016)\citenamefont{Sch\"{o}ne,
  Kr\"{u}ger, Gr\"{u}nwald, Stolz, Scheel, A{\ss}mann, Heck\"{o}tter, Thewes,
  Fr\"{o}hlich, and Bayer}}]{80}
\bibinfo{author}{\bibfnamefont{F.}~\bibnamefont{Sch\"{o}ne}},
  \bibinfo{author}{\bibfnamefont{S.~O.} \bibnamefont{Kr\"{u}ger}},
  \bibinfo{author}{\bibfnamefont{P.}~\bibnamefont{Gr\"{u}nwald}},
  \bibinfo{author}{\bibfnamefont{H.}~\bibnamefont{Stolz}},
  \bibinfo{author}{\bibfnamefont{S.}~\bibnamefont{Scheel}},
  \bibinfo{author}{\bibfnamefont{M.}~\bibnamefont{A{\ss}mann}},
  \bibinfo{author}{\bibfnamefont{J.}~\bibnamefont{Heck\"{o}tter}},
  \bibinfo{author}{\bibfnamefont{J.}~\bibnamefont{Thewes}},
  \bibinfo{author}{\bibfnamefont{D.}~\bibnamefont{Fr\"{o}hlich}},
  \bibnamefont{and} \bibinfo{author}{\bibfnamefont{M.}~\bibnamefont{Bayer}},
  \bibinfo{journal}{Phys. Rev. B} \textbf{\bibinfo{volume}{93}},
  \bibinfo{pages}{075203} (\bibinfo{year}{2016}).

\bibitem[{\citenamefont{Schweiner
  et~al.}(2016{\natexlab{a}})\citenamefont{Schweiner, Main, and Wunner}}]{75}
\bibinfo{author}{\bibfnamefont{F.}~\bibnamefont{Schweiner}},
  \bibinfo{author}{\bibfnamefont{J.}~\bibnamefont{Main}}, \bibnamefont{and}
  \bibinfo{author}{\bibfnamefont{G.}~\bibnamefont{Wunner}},
  \bibinfo{journal}{Phys. Rev. B} \textbf{\bibinfo{volume}{93}},
  \bibinfo{pages}{085203} (\bibinfo{year}{2016}{\natexlab{a}}).

\bibitem[{\citenamefont{Schweiner
  et~al.}(2016{\natexlab{b}})\citenamefont{Schweiner, Main, Feldmaier, Wunner,
  and Uihlein}}]{100}
\bibinfo{author}{\bibfnamefont{F.}~\bibnamefont{Schweiner}},
  \bibinfo{author}{\bibfnamefont{J.}~\bibnamefont{Main}},
  \bibinfo{author}{\bibfnamefont{M.}~\bibnamefont{Feldmaier}},
  \bibinfo{author}{\bibfnamefont{G.}~\bibnamefont{Wunner}}, \bibnamefont{and}
  \bibinfo{author}{\bibfnamefont{{\relax Ch}.}~\bibnamefont{Uihlein}},
  \bibinfo{journal}{Phys. Rev. B} \textbf{\bibinfo{volume}{93}},
  \bibinfo{pages}{195203} (\bibinfo{year}{2016}{\natexlab{b}}).

\bibitem[{\citenamefont{Zieli\'{n}ska-Raczy\'{n}ska
  et~al.}(2016{\natexlab{a}})\citenamefont{Zieli\'{n}ska-Raczy\'{n}ska,
  Czajkowski, and Ziemkiewicz}}]{74}
\bibinfo{author}{\bibfnamefont{S.}~\bibnamefont{Zieli\'{n}ska-Raczy\'{n}ska}},
  \bibinfo{author}{\bibfnamefont{G.}~\bibnamefont{Czajkowski}},
  \bibnamefont{and}
  \bibinfo{author}{\bibfnamefont{D.}~\bibnamefont{Ziemkiewicz}},
  \bibinfo{journal}{Phys. Rev. B} \textbf{\bibinfo{volume}{93}},
  \bibinfo{pages}{075206} (\bibinfo{year}{2016}{\natexlab{a}}).

\bibitem[{\citenamefont{Feldmaier et~al.}(2016)\citenamefont{Feldmaier, Main,
  Schweiner, Cartarius, and Wunner}}]{50}
\bibinfo{author}{\bibfnamefont{M.}~\bibnamefont{Feldmaier}},
  \bibinfo{author}{\bibfnamefont{J.}~\bibnamefont{Main}},
  \bibinfo{author}{\bibfnamefont{F.}~\bibnamefont{Schweiner}},
  \bibinfo{author}{\bibfnamefont{H.}~\bibnamefont{Cartarius}},
  \bibnamefont{and} \bibinfo{author}{\bibfnamefont{G.}~\bibnamefont{Wunner}},
  \bibinfo{journal}{J. Phys. B: At. Mol. Opt. Phys.}
  \textbf{\bibinfo{volume}{49}}, \bibinfo{pages}{144002}
  (\bibinfo{year}{2016}).

\bibitem[{\citenamefont{A{\ss}mann et~al.}(2016)\citenamefont{A{\ss}mann,
  Thewes, Fr\"{o}hlich, and Bayer}}]{QC}
\bibinfo{author}{\bibfnamefont{M.}~\bibnamefont{A{\ss}mann}},
  \bibinfo{author}{\bibfnamefont{J.}~\bibnamefont{Thewes}},
  \bibinfo{author}{\bibfnamefont{D.}~\bibnamefont{Fr\"{o}hlich}},
  \bibnamefont{and} \bibinfo{author}{\bibfnamefont{M.}~\bibnamefont{Bayer}},
  \bibinfo{journal}{Nature Mater.} \textbf{\bibinfo{volume}{15}},
  \bibinfo{pages}{741} (\bibinfo{year}{2016}).

\bibitem[{\citenamefont{Gr\"unwald et~al.}(2016)\citenamefont{Gr\"unwald,
  A{\ss}mann, Heck\"{o}tter, Fr\"{o}hlich, Bayer, Stolz, and Scheel}}]{76}
\bibinfo{author}{\bibfnamefont{P.}~\bibnamefont{Gr\"unwald}},
  \bibinfo{author}{\bibfnamefont{M.}~\bibnamefont{A{\ss}mann}},
  \bibinfo{author}{\bibfnamefont{J.}~\bibnamefont{Heck\"{o}tter}},
  \bibinfo{author}{\bibfnamefont{D.}~\bibnamefont{Fr\"{o}hlich}},
  \bibinfo{author}{\bibfnamefont{M.}~\bibnamefont{Bayer}},
  \bibinfo{author}{\bibfnamefont{H.}~\bibnamefont{Stolz}}, \bibnamefont{and}
  \bibinfo{author}{\bibfnamefont{S.}~\bibnamefont{Scheel}},
  \bibinfo{journal}{Phys. Rev. Lett.} \textbf{\bibinfo{volume}{117}},
  \bibinfo{pages}{133003} (\bibinfo{year}{2016}).

\bibitem[{\citenamefont{Zieli\'{n}ska-Raczy\'{n}ska
  et~al.}(2016{\natexlab{b}})\citenamefont{Zieli\'{n}ska-Raczy\'{n}ska,
  Ziemkiewicz, and Czajkowski}}]{77}
\bibinfo{author}{\bibfnamefont{S.}~\bibnamefont{Zieli\'{n}ska-Raczy\'{n}ska}},
  \bibinfo{author}{\bibfnamefont{D.}~\bibnamefont{Ziemkiewicz}},
  \bibnamefont{and}
  \bibinfo{author}{\bibfnamefont{G.}~\bibnamefont{Czajkowski}},
  \bibinfo{journal}{Phys. Rev. B} \textbf{\bibinfo{volume}{94}},
  \bibinfo{pages}{045205} (\bibinfo{year}{2016}{\natexlab{b}}).

\bibitem[{\citenamefont{Schweiner
  et~al.}(2016{\natexlab{c}})\citenamefont{Schweiner, Main, Wunner, and
  Uihlein}}]{150}
\bibinfo{author}{\bibfnamefont{F.}~\bibnamefont{Schweiner}},
  \bibinfo{author}{\bibfnamefont{J.}~\bibnamefont{Main}},
  \bibinfo{author}{\bibfnamefont{G.}~\bibnamefont{Wunner}}, \bibnamefont{and}
  \bibinfo{author}{\bibfnamefont{{\relax Ch}.}~\bibnamefont{Uihlein}},
  \bibinfo{journal}{Phys. Rev. B} \textbf{\bibinfo{volume}{94}},
  \bibinfo{pages}{115201} (\bibinfo{year}{2016}{\natexlab{c}}).

\bibitem[{\citenamefont{Schweiner et~al.}(2017)\citenamefont{Schweiner, Main,
  Wunner, Freitag, Heck\"{o}tter, Uihlein, A{\ss}mann, Fr\"{o}hlich, and
  Bayer}}]{125}
\bibinfo{author}{\bibfnamefont{F.}~\bibnamefont{Schweiner}},
  \bibinfo{author}{\bibfnamefont{J.}~\bibnamefont{Main}},
  \bibinfo{author}{\bibfnamefont{G.}~\bibnamefont{Wunner}},
  \bibinfo{author}{\bibfnamefont{M.}~\bibnamefont{Freitag}},
  \bibinfo{author}{\bibfnamefont{J.}~\bibnamefont{Heck\"{o}tter}},
  \bibinfo{author}{\bibfnamefont{{\relax Ch}.}~\bibnamefont{Uihlein}},
  \bibinfo{author}{\bibfnamefont{M.}~\bibnamefont{A{\ss}mann}},
  \bibinfo{author}{\bibfnamefont{D.}~\bibnamefont{Fr\"{o}hlich}},
  \bibnamefont{and} \bibinfo{author}{\bibfnamefont{M.}~\bibnamefont{Bayer}},
  \bibinfo{journal}{Phys. Rev. B}  (\bibinfo{year}{2017}), \bibinfo{note}{in
  press, arXiv:1609.04275}.

\bibitem[{\citenamefont{Luttinger}(1956)}]{25}
\bibinfo{author}{\bibfnamefont{J.}~\bibnamefont{Luttinger}},
  \bibinfo{journal}{Phys. Rev.} \textbf{\bibinfo{volume}{102}},
  \bibinfo{pages}{1030} (\bibinfo{year}{1956}).

\bibitem[{\citenamefont{Baldereschi and Lipari}(1973)}]{7_11}
\bibinfo{author}{\bibfnamefont{A.}~\bibnamefont{Baldereschi}} \bibnamefont{and}
  \bibinfo{author}{\bibfnamefont{N.~O.} \bibnamefont{Lipari}},
  \bibinfo{journal}{Phys. Rev. B} \textbf{\bibinfo{volume}{8}},
  \bibinfo{pages}{2697} (\bibinfo{year}{1973}).

\bibitem[{\citenamefont{Uihlein et~al.}(1981)\citenamefont{Uihlein,
  Fr\"{o}hlich, and Kenklies}}]{7}
\bibinfo{author}{\bibfnamefont{{\relax Ch}.}~\bibnamefont{Uihlein}},
  \bibinfo{author}{\bibfnamefont{D.}~\bibnamefont{Fr\"{o}hlich}},
  \bibnamefont{and} \bibinfo{author}{\bibfnamefont{R.}~\bibnamefont{Kenklies}},
  \bibinfo{journal}{Phys. Rev. B} \textbf{\bibinfo{volume}{23}},
  \bibinfo{pages}{2731} (\bibinfo{year}{1981}).

\bibitem[{\citenamefont{Schmelcher and Cederbaum}(1992)}]{90}
\bibinfo{author}{\bibfnamefont{P.}~\bibnamefont{Schmelcher}} \bibnamefont{and}
  \bibinfo{author}{\bibfnamefont{L.~S.} \bibnamefont{Cederbaum}},
  \bibinfo{journal}{Z. Phys. D} \textbf{\bibinfo{volume}{24}},
  \bibinfo{pages}{311} (\bibinfo{year}{1992}).

\bibitem[{\citenamefont{Schmelcher and Cederbaum}(1993)}]{91}
\bibinfo{author}{\bibfnamefont{P.}~\bibnamefont{Schmelcher}} \bibnamefont{and}
  \bibinfo{author}{\bibfnamefont{L.~S.} \bibnamefont{Cederbaum}},
  \bibinfo{journal}{Phys. Rev. A} \textbf{\bibinfo{volume}{47}},
  \bibinfo{pages}{2634} (\bibinfo{year}{1993}).

\bibitem[{\citenamefont{Altarelli and Lipari}(1973)}]{34}
\bibinfo{author}{\bibfnamefont{M.}~\bibnamefont{Altarelli}} \bibnamefont{and}
  \bibinfo{author}{\bibfnamefont{N.~O.} \bibnamefont{Lipari}},
  \bibinfo{journal}{Phys. Rev. B} \textbf{\bibinfo{volume}{7}},
  \bibinfo{pages}{3798} (\bibinfo{year}{1973}).

\bibitem[{\citenamefont{Altarelli and Lipari}(1974)}]{33}
\bibinfo{author}{\bibfnamefont{M.}~\bibnamefont{Altarelli}} \bibnamefont{and}
  \bibinfo{author}{\bibfnamefont{N.~O.} \bibnamefont{Lipari}},
  \bibinfo{journal}{Phys. Rev. B} \textbf{\bibinfo{volume}{9}},
  \bibinfo{pages}{1733} (\bibinfo{year}{1974}).

\bibitem[{\citenamefont{Chen et~al.}(1987)\citenamefont{Chen, Gil, Mathieu, and
  Lascaray}}]{39}
\bibinfo{author}{\bibfnamefont{Y.}~\bibnamefont{Chen}},
  \bibinfo{author}{\bibfnamefont{B.}~\bibnamefont{Gil}},
  \bibinfo{author}{\bibfnamefont{H.}~\bibnamefont{Mathieu}}, \bibnamefont{and}
  \bibinfo{author}{\bibfnamefont{J.~P.} \bibnamefont{Lascaray}},
  \bibinfo{journal}{Phys. Rev. B} \textbf{\bibinfo{volume}{36}},
  \bibinfo{pages}{1510} (\bibinfo{year}{1987}).

\bibitem[{\citenamefont{Knox}(1963)}]{TOE}
\bibinfo{author}{\bibfnamefont{R.}~\bibnamefont{Knox}},
  \emph{\bibinfo{title}{Theory of excitons}}, vol.~\bibinfo{volume}{5} of
  \emph{\bibinfo{series}{Solid State Physics Supplement}}
  (\bibinfo{publisher}{Academic}, \bibinfo{address}{New York},
  \bibinfo{year}{1963}).

\bibitem[{\citenamefont{Broeckx}(1991)}]{44}
\bibinfo{author}{\bibfnamefont{J.}~\bibnamefont{Broeckx}},
  \bibinfo{journal}{Phys. Rev. B} \textbf{\bibinfo{volume}{43}},
  \bibinfo{pages}{9643} (\bibinfo{year}{1991}).

\bibitem[{\citenamefont{Suzuki and Hensel}(1974)}]{44_12}
\bibinfo{author}{\bibfnamefont{K.}~\bibnamefont{Suzuki}} \bibnamefont{and}
  \bibinfo{author}{\bibfnamefont{J.~C.} \bibnamefont{Hensel}},
  \bibinfo{journal}{Phys. Rev. B} \textbf{\bibinfo{volume}{9}},
  \bibinfo{pages}{4184} (\bibinfo{year}{1974}).

\bibitem[{\citenamefont{Messiah}(1969)}]{Messiah2}
\bibinfo{author}{\bibfnamefont{A.}~\bibnamefont{Messiah}},
  \emph{\bibinfo{title}{Quantum Mechanics}}, vol.~\bibinfo{volume}{2}
  (\bibinfo{publisher}{North-Holland}, \bibinfo{address}{Amsterdam},
  \bibinfo{year}{1969}).

\bibitem[{\citenamefont{Caprio et~al.}(2012)\citenamefont{Caprio, Maris, and
  Vary}}]{S1}
\bibinfo{author}{\bibfnamefont{M.~A.} \bibnamefont{Caprio}},
  \bibinfo{author}{\bibfnamefont{P.}~\bibnamefont{Maris}}, \bibnamefont{and}
  \bibinfo{author}{\bibfnamefont{J.~P.} \bibnamefont{Vary}},
  \bibinfo{journal}{Phys. Rev. C} \textbf{\bibinfo{volume}{86}},
  \bibinfo{pages}{034312} (\bibinfo{year}{2012}).

\bibitem[{\citenamefont{Edmonds}(1960)}]{ED}
\bibinfo{author}{\bibfnamefont{A.}~\bibnamefont{Edmonds}},
  \emph{\bibinfo{title}{Angular momentum in quantum mechanics}}
  (\bibinfo{publisher}{Princeton University Press},
  \bibinfo{address}{Princeton}, \bibinfo{year}{1960}).

\bibitem[{\citenamefont{Anderson et~al.}(1999)\citenamefont{Anderson, Bai,
  Bischof, Blackford, Demmel, Dongarra, Croz, Greenbaum, Hammarling, McKenney
  et~al.}}]{Lapack}
\bibinfo{author}{\bibfnamefont{E.}~\bibnamefont{Anderson}},
  \bibinfo{author}{\bibfnamefont{Z.}~\bibnamefont{Bai}},
  \bibinfo{author}{\bibfnamefont{C.}~\bibnamefont{Bischof}},
  \bibinfo{author}{\bibfnamefont{S.}~\bibnamefont{Blackford}},
  \bibinfo{author}{\bibfnamefont{J.}~\bibnamefont{Demmel}},
  \bibinfo{author}{\bibfnamefont{J.}~\bibnamefont{Dongarra}},
  \bibinfo{author}{\bibfnamefont{J.~D.} \bibnamefont{Croz}},
  \bibinfo{author}{\bibfnamefont{A.}~\bibnamefont{Greenbaum}},
  \bibinfo{author}{\bibfnamefont{S.}~\bibnamefont{Hammarling}},
  \bibinfo{author}{\bibfnamefont{A.}~\bibnamefont{McKenney}},
  \bibnamefont{et~al.}, \emph{\bibinfo{title}{{LAPACK} Users' Guide}}
  (\bibinfo{publisher}{Society for Industrial and Applied Mathematics},
  \bibinfo{address}{Philadelphia, PA}, \bibinfo{year}{1999}),
  \bibinfo{edition}{3rd} ed.

\bibitem[{\citenamefont{Brody et~al.}(1981{\natexlab{b}})\citenamefont{Brody,
  Flores, French, Mello, Pandey, and Wong}}]{QC_16}
\bibinfo{author}{\bibfnamefont{T.~A.} \bibnamefont{Brody}},
  \bibinfo{author}{\bibfnamefont{J.}~\bibnamefont{Flores}},
  \bibinfo{author}{\bibfnamefont{J.~B.} \bibnamefont{French}},
  \bibinfo{author}{\bibfnamefont{P.~A.} \bibnamefont{Mello}},
  \bibinfo{author}{\bibfnamefont{A.}~\bibnamefont{Pandey}}, \bibnamefont{and}
  \bibinfo{author}{\bibfnamefont{S.~S.~M.} \bibnamefont{Wong}},
  \bibinfo{journal}{Rev. Mod. Phys.} \textbf{\bibinfo{volume}{53}},
  \bibinfo{pages}{385} (\bibinfo{year}{1981}{\natexlab{b}}).

\bibitem[{\citenamefont{Grosa et~al.}(2013)\citenamefont{Grosa, Legranda,
  Mortessagnea, Richalotb, and Selemanib}}]{GUE2}
\bibinfo{author}{\bibfnamefont{J.-B.} \bibnamefont{Grosa}},
  \bibinfo{author}{\bibfnamefont{O.}~\bibnamefont{Legranda}},
  \bibinfo{author}{\bibfnamefont{F.}~\bibnamefont{Mortessagnea}},
  \bibinfo{author}{\bibfnamefont{E.}~\bibnamefont{Richalotb}},
  \bibnamefont{and}
  \bibinfo{author}{\bibfnamefont{K.}~\bibnamefont{Selemanib}},
  \bibinfo{journal}{Wave Motion} \textbf{\bibinfo{volume}{51}},
  \bibinfo{pages}{664} (\bibinfo{year}{2013}).

\end{thebibliography}

\end{document}